\documentclass[fleqn,usenatbib]{mnras}

\usepackage{newtxtext,newtxmath}

\usepackage[T1]{fontenc}

\DeclareRobustCommand{\VAN}[3]{#2}
\let\VANthebibliography\thebibliography
\def\thebibliography{\DeclareRobustCommand{\VAN}[3]{##3}\VANthebibliography}

\usepackage{graphicx}	
\graphicspath{{figures/}{./}}
\usepackage{amsmath}	

\title[Unbinned Likelihood]{Unbinned Likelihood Analysis for X-ray Polarization}

\author[Gonz{\'a}lez-Caniulef, Caiazzo \& Heyl]{
Denis Gonz{\'a}lez-Caniulef\thanks{denis.gonzalez-caniulef@irap.omp.eu, dgonzalez@phas.ubc.ca; CITA National Fellow}$^1$,
Ilaria Caiazzo\thanks{email: ilariac@caltech.edu; Sherman Fairchild Fellow}$^2$,
Jeremy Heyl\thanks{email: heyl@phas.ubca.ca}$^1$
\\
$^{1}$Department of Physics and Astronomy, University of British Columbia, Vancouver, BC V6T 1Z1, Canada \\
$^{2}$TAPIR, Walter Burke Institute for Theoretical Physics, Mail Code 350-17, Caltech, Pasadena, CA 91125, USA
}

\date{Accepted XXX. Received YYY; in original form ZZZ}

\pubyear{2021}

\begin{document}
\label{firstpage}
\pagerange{\pageref{firstpage}--\pageref{lastpage}}
\maketitle
\begin{abstract}
We present a systematic study of the unbinned, photon-by-photon likelihood technique which can be used as an alternative method to analyse phase-dependent, X-ray spectro-polarimetric observations obtained with IXPE and other photo-electric polarimeters. 
We apply the unbinned technique to models of the luminous X-ray pulsar Hercules X-1, for which we produce simulated observations using \texttt{ixpeobssim} package.
We consider minimal knowledge about the actual physical process responsible for the polarized emission from the accreting pulsar and assume that the observed phase-dependent polarization angle can be described by the rotating vector model.  
Using the unbinned technique, the detector’s modulation factor, and the polarization information alone, we found that both the rotating vector model and the underlying spectro-polarimetry model can reconstruct equally well the geometric configuration angles of the accreting pulsar. 
However, the measured polarization fraction becomes biased with respect to underlying model unless the  energy dispersion and effective area of the detector are also taken into account. \textcolor{black}{To this end, we present an  energy-dispersed likelihood estimator that is proved to be unbiased.}
For different analyses, we obtain posterior distributions from multiple \texttt{ixpeobssim} realizations and show that the unbinned technique yields  $\sim 10\%$ smaller error bars than the binned technique. We also discuss alternative sources, such as magnetars, in which the unbinned technique and the rotating vector model might be applied. 
\end{abstract}




\section{Introduction}
\label{sec:introduction}

The Imaging X-ray Polarimetry Explorer mission \citep[IXPE,][]{Weisskopf16} was successfully launched on December 2021, opening a new window to study X-ray sources and physical processes in extreme astrophysical environments. IXPE consists of three identical telescopes, each of them carrying an independent Gas Pixel Detector (GPD) polarimeter instrument ($2-8$ keV range), whose technology is based on the photoelectric effect \citep{Costa2001,Bellazzini2007}.  A beam of radiation from a source results in multiple photoelectron tracks \textcolor{black}{that are individually detected}  in the GPD, from which is possible to  reconstruct,  \textcolor{black}{via the photoelectric scattering cross section}, the initial energy and polarization direction of single photons. The rich polarimetric information contained in IXPE observations naturally stimulates the development of new data analysis techniques.

The analysis of X-ray observations often relies on the binning technique, in which a list of photons is pre-processed by grouping them into bins (e.g. in energy bins, phase bins, etc.) and then characterized by the counts in each interval. 
By binning a photon list, information is inevitably lost, whose impact on the analysis also depends on the type of binning choice.
This loss can be mitigated by a large sampling of the data or by adapting the binning according the different science cases, which usually do not have an unique criteria. 

However, this becomes a major issue when the analysis has to deal with scarce data, e.g. low count rates, large background component, etc.
An alternative method to the binned analysis is the photon-by-photon likelihood analysis, also known as unbinned likelihood method. This method has been already routinely used, for example in gamma-ray astronomy, and it has been also discussed for X-ray polarimetry \citep[see e.g.][]{2021AJ....162..134M}, as well as in the context of photoelectron track analysis (using convolutional neural networks), aiming to improve the sensitivity (modulation factor) of GPDs \citep{Peirson2021a, Peirson2021b}. Our goal is to use the unbinned technique to compare a model against photon-by-photon events\footnote{A Jupyter Notebook is available in the link below:\\ \url{https://github.com/UBC-Astrophysics/Unbinned-Xray-polarimetry}} to achieve a higher sensitivity beyond the chi-square analysis of binned polarimetry data \citep[see e.g.][]{Kislat2015}, \textcolor{black}{either from IXPE observations or other X-ray polarimeters.}\footnote{\textcolor{black}{For example:  eXTP \citep{Zhang16}, XPP  \citep{Jahoda19}, PolarLight \citep{Feng19}, X-Calibur \citep{Beilicke2014},   GOSoX \citep{Marshall2021},  POLIX \citep{Vadawale2010}, PoGO+ \citep{Chauvin2018}, LAMP \citep{She2015}, etc.}}
Besides extracting the maximum amount of information from the data, the unbinned technique can be implemented quite economically even when a variety of instrumental response functions need to be considered in the analysis.

We here study the unbinned technique and apply it to simulations of IXPE observations of  the accreting X-ray pulsar Hercules X-1 (hereafter Her X-1), which is one of the main targets in IXPE's long term plan.  The simulations are generated using \texttt{ixpeobssim} package \citep[see also \citealt{Pesce2019}]{Baldini2022}, while the model for the polarized X-ray emission from the X-ray pulsar is taken  from  \citet{2021MNRAS.501..109C} and \citet{2021MNRAS.501..129C}.
We perform various tests \textcolor{black}{  and statistical assessments from multiple realizations for both} the binned and unbinned technique,  comparing the simulated data with the rotating vector model (RVM) and the underlying model.  \textcolor{black}{ Finally, we also present an unbiased, energy-dispersed likelihood estimator and show} an example to implement it in Python.

The paper is organized as follows. In \S~2,  we present the method and derivation of the unbinned technique. In \S~3, we discuss the results of our analysis of Her X-1 simulated data using the unbinned and binned technique. Conclusions are presented  in \S~4.

\section{The Method}
\label{sec:method}

The IXPE observatory consists of three X-ray telescopes mounted in parallel, each feeding radiation into a GPD \citep[for recent discussions see e.g.][]{Soffitta2021,Baldini2021}. The X-ray photons are absorbed by atoms in the gas (dimethyl ether) and ionize an electron from the K-shell.  The ionization is typically well above the threshold of the particular atom, so the cross section decreases with increasing photon energy.  Furthermore, the outgoing electron is approximately in a well-defined momentum state; the direction of the momentum correlates with the photon electric field with a $\cos^2 \psi$ dependence.  

The magnitude of the correlation for fully linearly polarized radiation is known as the modulation factor.  By modelling the high-energy photo-electron track \citep{2017SPIE10397E..0FS}, the following properties of the incoming photon can be determined: energy, arrival time, electron direction and position on the sky.  As the energy of the photon increases, so does the momentum of the photo-electron, which results in a longer and more easily measured track, and therefore the modulation factor for IXPE increases with increasing energy from 15\% at 2~keV to 60\% at 8~keV \citep{2021arXiv211201269W}.  

We can build the likelihood function focusing only on the polarization signal, which is a function of photon angle, energy and time, or by including the spectrum, which is just a function of energy and time, in the model as well  \citep[see also][]{Kislat2015,2021AJ....162..134M}. Let $p_0(E,t)$ be polarization degree of the model, $\mu(E)$ be the modulation factor of the instrument and $\psi_0(E,t)$ be the polarization angle of the model; the first component of the likelihood function can therefore be written as:
$$
f(\psi) = \frac{1}{2\pi} \left [  1 +  \mu p_0 \left( 2 \cos^2 (\psi-\psi_0)-1 \right)  \right]
$$
where $\psi$ is the angle of a particular photon and the energy and time variables are suppressed.  This expression results from the definition of the modulation factor, the differential scattering cross-section, and the normalization $\int f(\psi) d \psi=1$, which is constant with respect to the expected degree of polarization.  We can reformulate it as
$$
f(\psi) = \frac{1}{2\pi} \left  [ 1 + \mu p_0 \left [ \cos2\psi \cos2\psi_0 + \sin2\psi\sin2\psi_0 \right] \right ],
$$
and, if we define $Q_m=p_0\cos2\psi_0$ and $U_m=p_0\sin2\psi_0$ for the model, and $Q_\gamma= \cos2\psi$ and $U_\gamma=\sin2\psi$ for the photon, we have
$$
f(\psi) =  \frac{1}{2\pi} \left  [1 + \mu \left [ Q_\gamma Q_m + U_\gamma U_m \right] \right ]
$$
If we want to include the spectrum $I(E,t)$ in the model as well, we can write
$$
f_i =  \frac{1}{2\pi} \left [1 + \left (\mu \left [ Q_\gamma Q_m + U_\gamma U_m \right] \right ) \right ]I =  \frac{1}{2\pi} {\bf S}_\gamma M {\bf S}
$$
where ${\bf S}$ are the Stokes parameters predicted by the model for the energy and arrival time of the photon, ${\bf S_\gamma}$ are the Stokes parameters of the detected photon and $M$ is the modulation matrix for the energy of the photon. $M$ is a property of the instrument; for IXPE it can be written as
$$
M = \left[ \begin{array}{cccc}
1 &            0                 &              0               & 0 \\
0 & {\mu(E)}         &              0               & 0\\
0 &            0                 & {\mu(E)}          & 0 \\
0 &            0                 &              0               & 0 
\end{array} \right ],
$$
where $\mu(E)$ is the modulation factor as a function of energy.  The last diagonal term is zero because IXPE does not detect circular polarization.   The matrix can have off-diagonal terms if there is mixing between polarization states (in the second to last column) or if the instrument exhibits spurious polarization  \citep[in the first column; for a recent study see e.g.][and references therein]{Rankin22}. In the case of IXPE, the effects of spurious polarization are accounted for by adding a term to the values of $Q_\gamma$ and $U_\gamma$.

The photon polarization vector is
$$
{\bf S}_{\gamma} = \left[ \begin{array}{c}
1 \\
\cos2\psi \cos2\chi \\
\sin2\psi \cos2\chi \\
\sin2\chi
\end{array} \right ].
$$
For IXPE, $\chi=0$ as the instrument does not detect circular polarization. The total logarithmic likelihood of the data given the model is
$$
\log L = \sum \log f_i - N_\textrm{pred}\,,
$$
where $N_\textrm{pred}$ is the number of photons predicted by the model.  For the more general situation where the instrument is also capable of detecting circular polarization we have
$$
M = \left[ \begin{array}{cccc}
1 &            0                 &              0               & 0 \\
0 & {\mu(E)}         &              0               & 0\\
0 &            0                 & {\mu(E)}         & 0 \\
0 &            0                 &              0               & {\mu(E)}
\end{array} \right ]
$$
assuming that the three modulation factors are equal.  

\section{Results}
\label{sec:results} 

As a first check, we simulate \textcolor{black}{ a simple bin of radiation consisting of} 1,000 fully polarized photons ($Q/I=\frac{1}{2}\sqrt{2}$ and $U/I=\frac{1}{2}\sqrt{2}$) assuming a modulation factor of unity. We recover the \textcolor{black}{total} value of $Q/I$ and $U/I$ using both the \textcolor{black}{maximum} likelihood estimator and the \citet{Kislat2015} estimator, twice the weighted (inversely with modulation factor) mean of $Q_\gamma$ and $U_\gamma$. \textcolor{black}{By repeating the test over 100,000 realisations},  we find that both estimators are unbiased and the unbinned estimator is about 32\% more efficient than the \citet{Kislat2015} estimator \textcolor{black}{ from the standard deviation of $Q/I$ and $U/I$ , which is 2.93\% versus 3.86\%}.  The latter value agrees with the \citet{Kislat2015} estimate of the variance. The increased efficiency means that for fully polarized radiation one can achieve the same precision with a 40\% shorter exposure time.  For unpolarized radiation, the two techniques perform equally well with a standard deviation of 4.4\% following the  \citet{Kislat2015} formula. Although the unbinned technique is  more efficient for polarized sources, the key advantages come into play when one expects the polarization to vary with time according to an underlying model and more subtly when the polarization varies with energy.

\begin{figure}
\includegraphics[width=8.5cm]{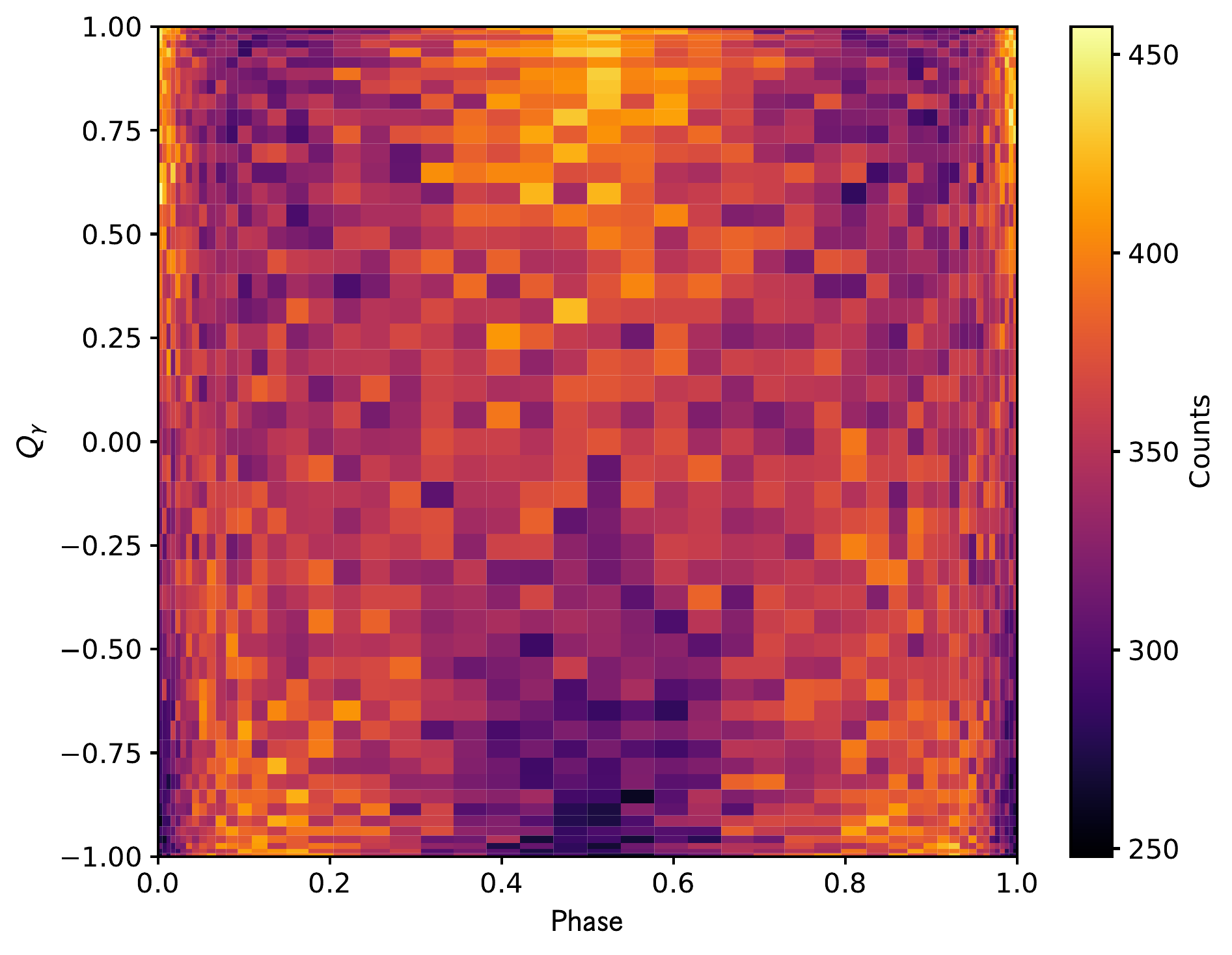}	
\includegraphics[width=8.5cm]{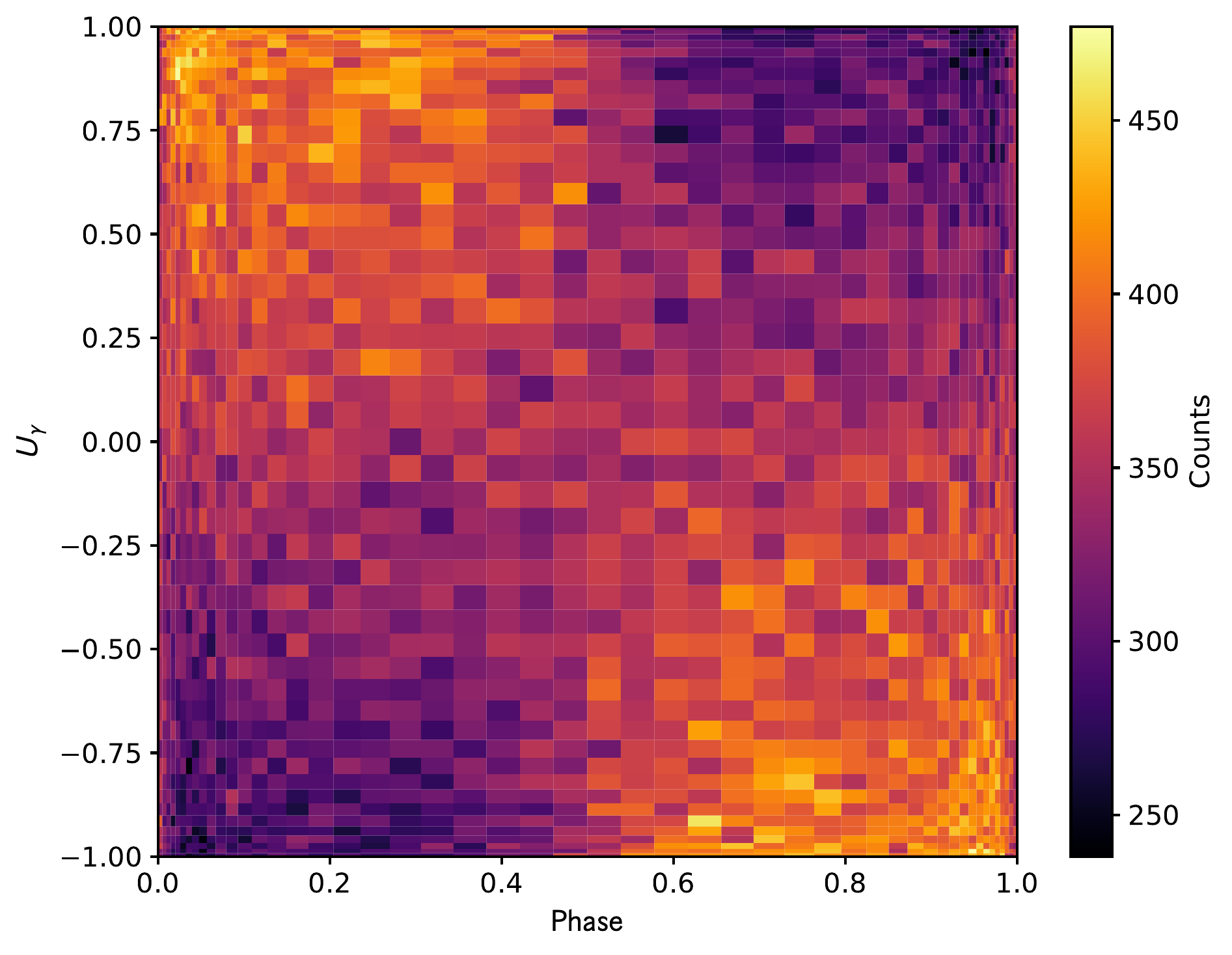}
\caption{\textcolor{black}{Stokes parameters ${\bf S_\gamma}$  simulated with \texttt{ixpeobssim} package for a  $100$\,ks observation of Her X-1 \citep[using the underlying model of][]{2021MNRAS.501..129C}. The upper and lower panels show 2-dimensional, equi-populated histograms for  the raw $Q_\gamma$ and $U_\gamma$ associated to each photon versus the rotational phase of the pulsar ($50\times50$ mesh equi-populated in phase and equi-spaced in photon angle $\psi$). Events are in the $2-8$ keV band and normalized to lie in the Stokes parameter range $[-1,1]$.}}
\label{fig:hist2dQU}
\end{figure}

\begin{figure}
\includegraphics[width=\columnwidth]{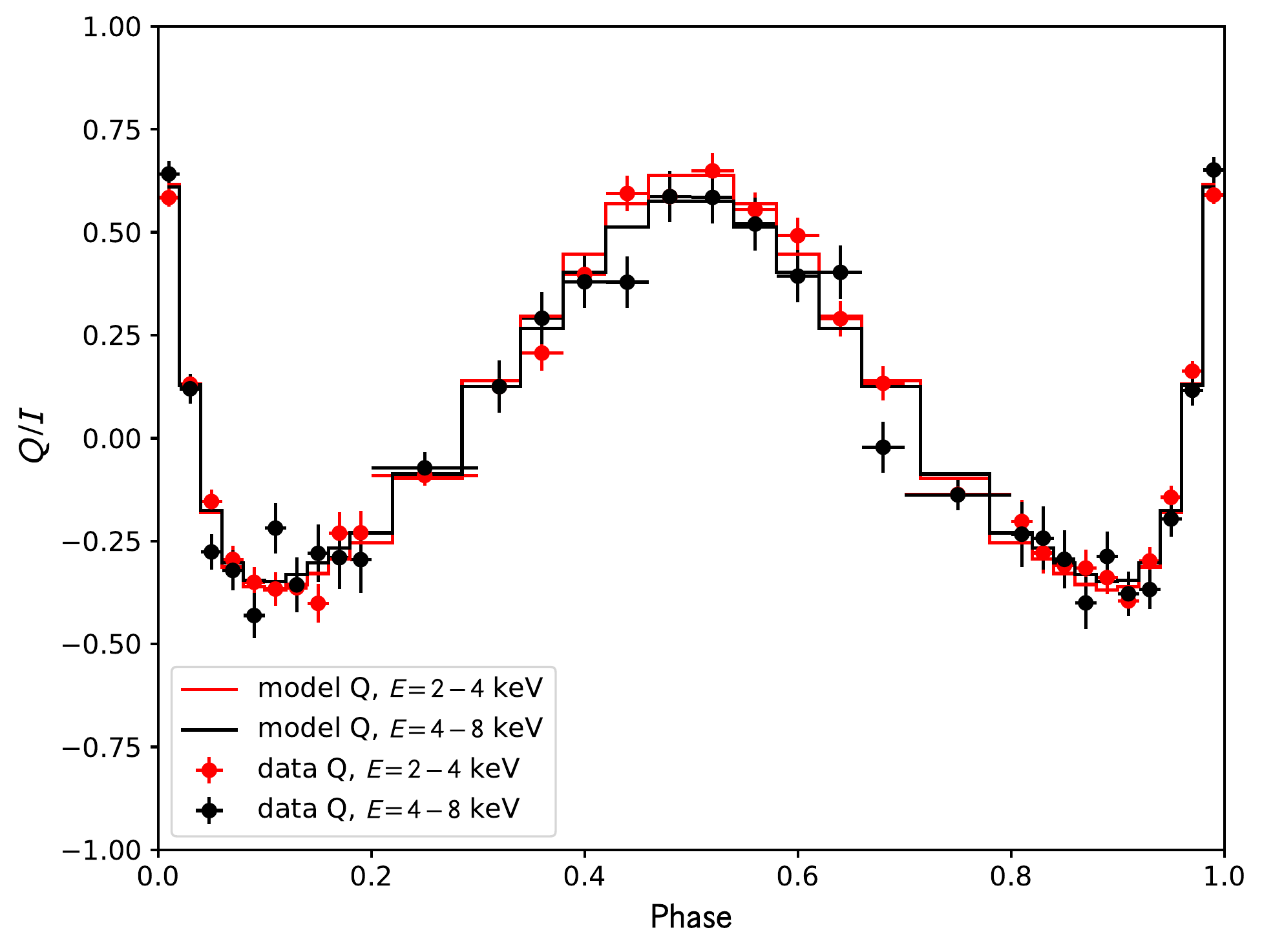}
\includegraphics[width=\columnwidth]{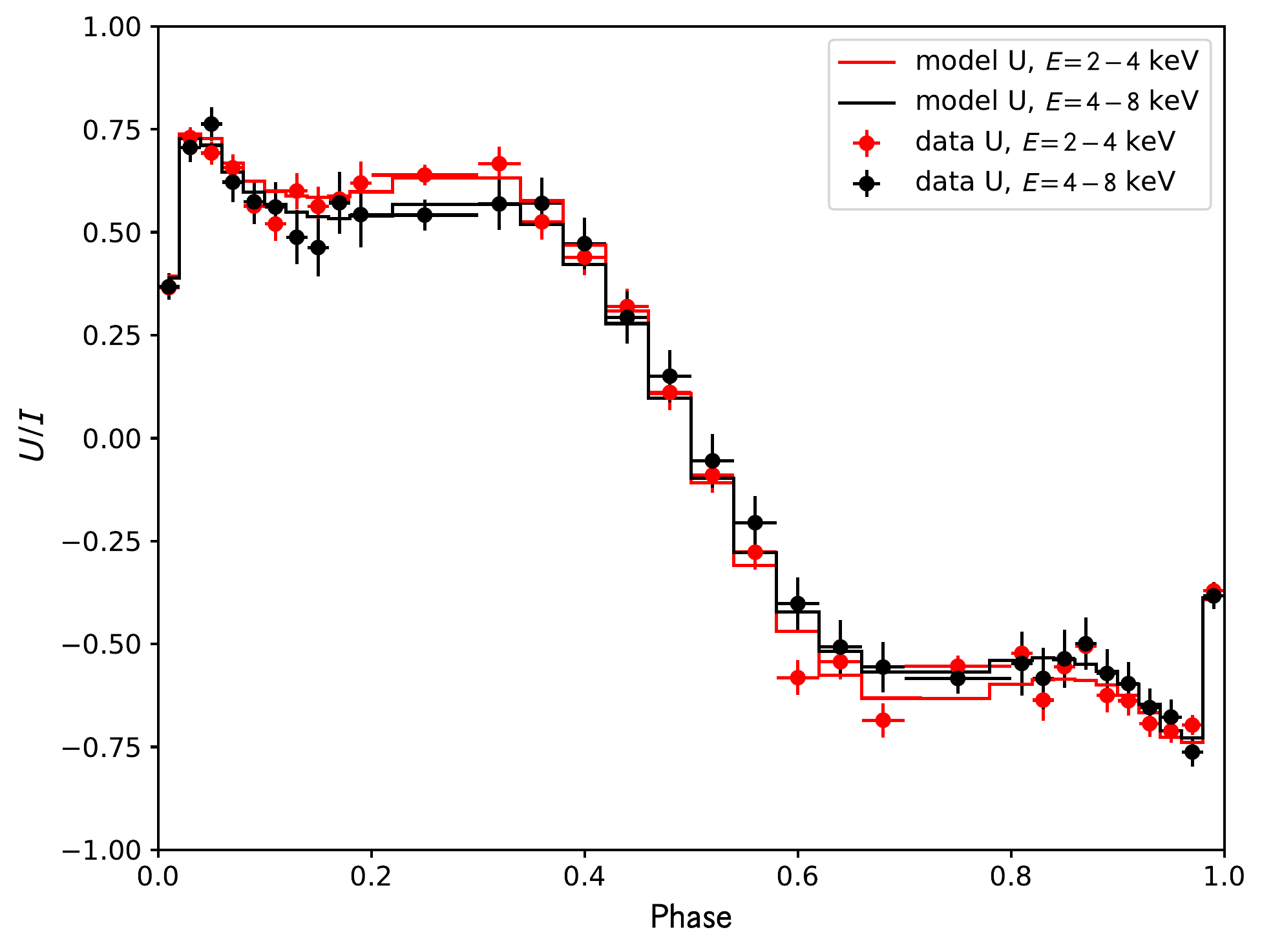}
\caption{Binned Stokes parameters in phase and energy for  Her X-1. The points with error bars correspond to a $100$\,ks simulated observation generated with \texttt{ixpeobssim} (same as in Figure\,\ref{fig:hist2dQU}). The solid lines correspond to the binned, underlying model of \citet{2021MNRAS.501..129C}}
\label{fig:binned_stokes}
\end{figure}

\subsection{\texttt{ixpeobssim} simulation}

We test both the binned and  unbinned likelihood technique using the models for X-ray emission from X-ray pulsars from \citet{2021MNRAS.501..109C} and \citet{2021MNRAS.501..129C} by varying the geometric parameters of the pulsar. \textcolor{black}{Both the polarization degree  and polarization angle in the model depend on the line-of-sight with respect to the magnetic axis of the pulsar. However, while the polarization degree is energy-dependent, the polarization angle is energy-independent in the IXPE band.} 

The pulsar model is tabulated in a file containing the spectra and Stokes parameters for $10^3$ energy bins logarithmically spaced between 0.5 to 77.9 keV,  and a mesh of $10^5$ line-of-sight angles relative to the magnetic axis.
Since the Stokes parameters for the tabulated models are defined relative to the direction of the magnetic field axis  of the pulsar, which is also  axisymmetric, only $Q/I$ is relevant, while $U/I=0$ in that frame. The model is generalized  to account for the pulsar rotation considering an arbitrary line-of-sight  angle $\alpha$ and  magnetic axis  angle $\beta$ with respect  to the spin axis, with  the Stokes $Q$ and $U$ redefined relative to the spin axis as well
\citep[see Appendix C in][]{2021MNRAS.501..109C}. 
\textcolor{black}{The model spectrum  is normalized according to the spectral flux from {\it NuSTAR} observations \citep{Wolff2016}, extrapolated to 1\,keV i.e., $F_{1-12 \,\mathrm{keV}}= 6\times 10^{-9} \mathrm{erg\, cm^{-2} \, s^{-1}}$}, considering a neutral hydrogen column density $N_\mathrm{H}=1.7\times10^{20}\,\mathrm{cm}^{-2}$ \citep{Furst2013}.

We generate periodic point source simulations of Her X-1 using  \texttt{ixpeobssim} package (version 18.0.0).
In the following, all simulations are run for 100-ks IXPE observations \textcolor{black}{ ($1.2\times10^6$ total events)}, setting the pulsar configuration angles $\alpha=52^{\circ}$ and $\beta=42^{\circ}$.
No background or spurious modulation are included  in the simulations.
For the pulsar ephemeris, we account only for the rotational frequency of the pulsar $f_0=0.806~\mathrm{Hz}$, while other parameters such as the frequency derivative and orbital period of the binary, for simplicity, are ignored.

We generate a photon list (stored in FITS files) from the main \texttt{ixpeobssim} Monte Carlo simulation,  accounting for the  instrument response functions of each IXPE detector unit. \textcolor{black}{The resulting photon list is then phase-folded, and the Stokes parameters ${\bf S_\gamma}$  of each event are directly used in the unbinned likelihood analysis as discussed in Sec.\,\ref{sec:method}. In order to visualize the simulated events, we found that  2-dimensional, equi-populated histograms of the photon Stokes $Q_\gamma$ and $U_\gamma$, as shown in Figure \ref{fig:hist2dQU}, already reveal the main polarization features of the underlying model.  On the other hand, in order to compute simultaneously the binned analysis, the events are binned}  considering two energy bands  (2$-$4 and 4$-$8~keV) and  
\textcolor{black}{ thirty-two phase bins with ten bins evenly spaced in phase from 0.3 to 0.7, twenty from 0.8 to 0.2, and two extra bins between the range's edges}.
Figure\,\ref{fig:binned_stokes}  shows the resulting binned  Stokes data  $Q/I$ and $U/I$ for a single \texttt{ixpeobssim} realization, as well as  the underlying model.

\subsection{Binned vs Unbinned analysis: No energy dispersion}
We first perform the polarimetry analysis comparing the simulated data to the models, taking into account only the detector's modulation factor. The analysis that includes the detector's energy dispersion is discussed in Section \ref{sec:energy_dispersion}. \textcolor{black}{In order to test both the binned and unbinned techniques,  we restrict the analysis to energies in the $2-4$\,keV range, where IXPE reach the maximum sensitivity to extract polarization information from the source.}

\subsubsection{Rotating vector model}
\label{sec:rvm}
Modelling the accretion process, coupled with the transport of polarized radiation, in a strongly magnetized NS is highly non-trivial \citep[see e.g.][and references therein]{2021MNRAS.501..109C}. Therefore,  we first analyse the data assuming that the underlying physical mechanism for the polarized emission from Her X-1 is unknown.   However, we do assume that the accretion is funneled into the magnetic poles through an axisymmetric magnetic field and that the radiation is emitted in the X-mode or O-mode \citep[see e.g.][]{Meszaros92}. Then, as a first approximation, the polarization angle of the radiation is determined by the direction of the pulsar magnetic axis relative to the spin axis, which can be modeled using the RVM \citep{Radhakrishnan1969}. The main idea is to reconstruct the geometric configuration of the pulsar using the information contained only in the  phase-dependent polarization angle, leaving the polarization degree and  spectral flux un-modelled. 

Similarly, the RVM might be also applied to other magnetized sources such as magnetars. As predicted by QED, under strong magnetic fields the vacuum becomes birefringent \citep[][]{Heisenberg1936,Weisskopf1936,Schwinger1951}. In the context of magnetars, this means that the radiation propagates in the magnetosphere readapting  the X- and O-mode to the local magnetic field. The main effect  is that the polarization angle detected by a distant observer reflects the magnetic field direction far from the NS's surface, where the field across the plane perpendicular to the line-of-sight is nearly uniform  \citep{Heyl2000, Heyl2002}. If the topology of the magnetic field is dominated by a dipolar component, without a strong toroidal component or twist \citep{Thompson02}, then the polarization angle is determined by the direction of magnetic axis with respect to the spin axis\footnote{\textcolor{black}{ This can be generally true even when vacuum birefringence is not operating as long as the topology of the magnetic field is nearly axisymmetric. In such case, the surface map of the Stokes parameters (for surface photons unaffected by resonant cyclotron scattering) averages the direction of the polarization angle also either along or perpendicular to the magnetic axis.}}. In this context, the RVM can be applied to analyse the phase-dependent polarization angle in the lower energy range of the soft X-ray spectrum, where X-ray photons have a low optical depth to resonant cyclotron scattering by charged particles streaming in the magnetar's magnetosphere \citep[][for recent reviews see also  \citealt{Turolla15, Kaspi17}]{Thompson02}. 

\begin{figure}
\centering
\includegraphics[width=8cm]{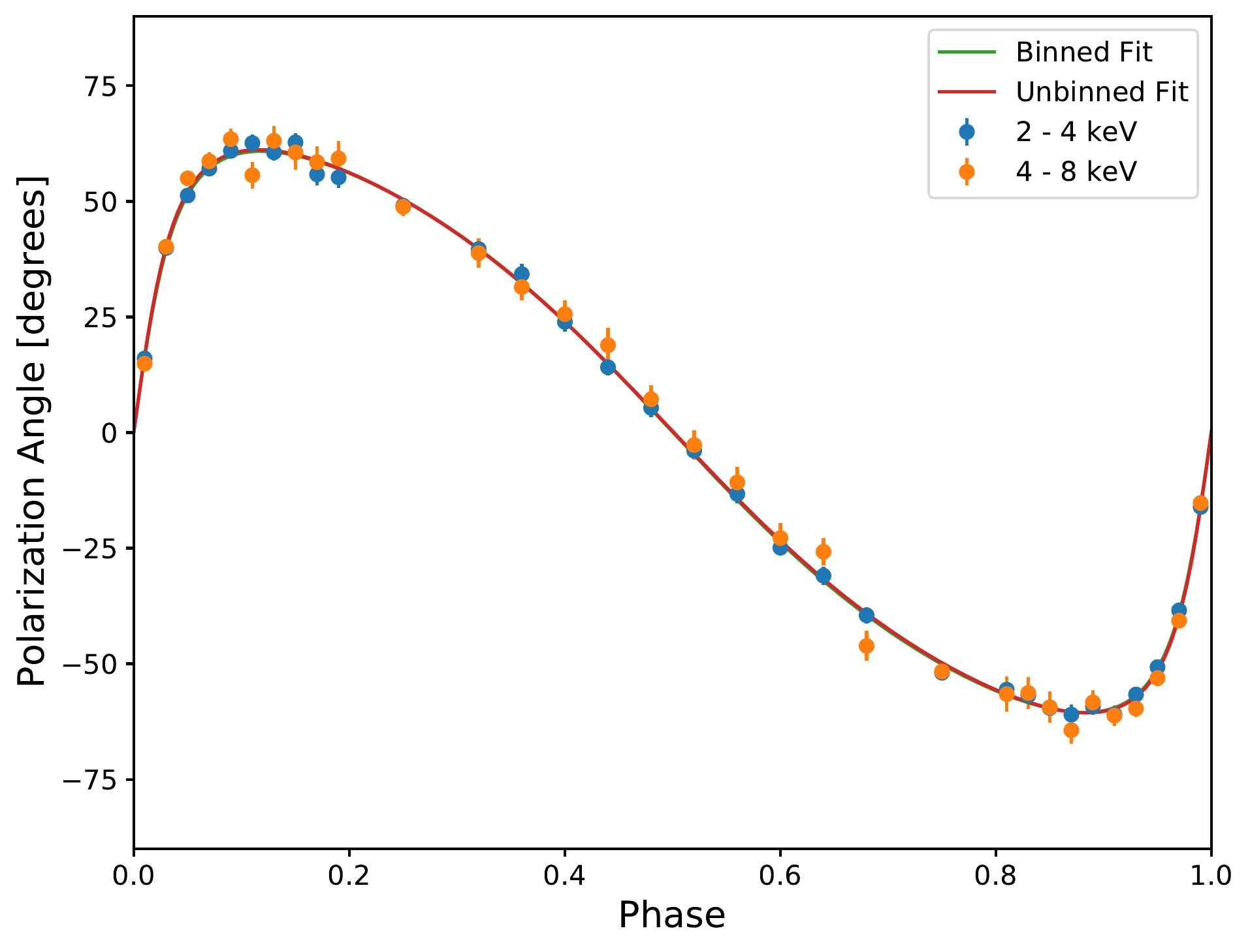}
\caption{The measured polarization angle from a simulated 100-ks observations of Her X-1 using the model of \citet{2021MNRAS.501..129C}.  The curves show the RVM model fitted for both the binned data from 2$-$4~keV and unbinned technique for photons identified with energies from 2$-$4~keV.}
    \label{fig:geomfit}
\end{figure}

\begin{figure}
\centering
\includegraphics[width=\columnwidth]{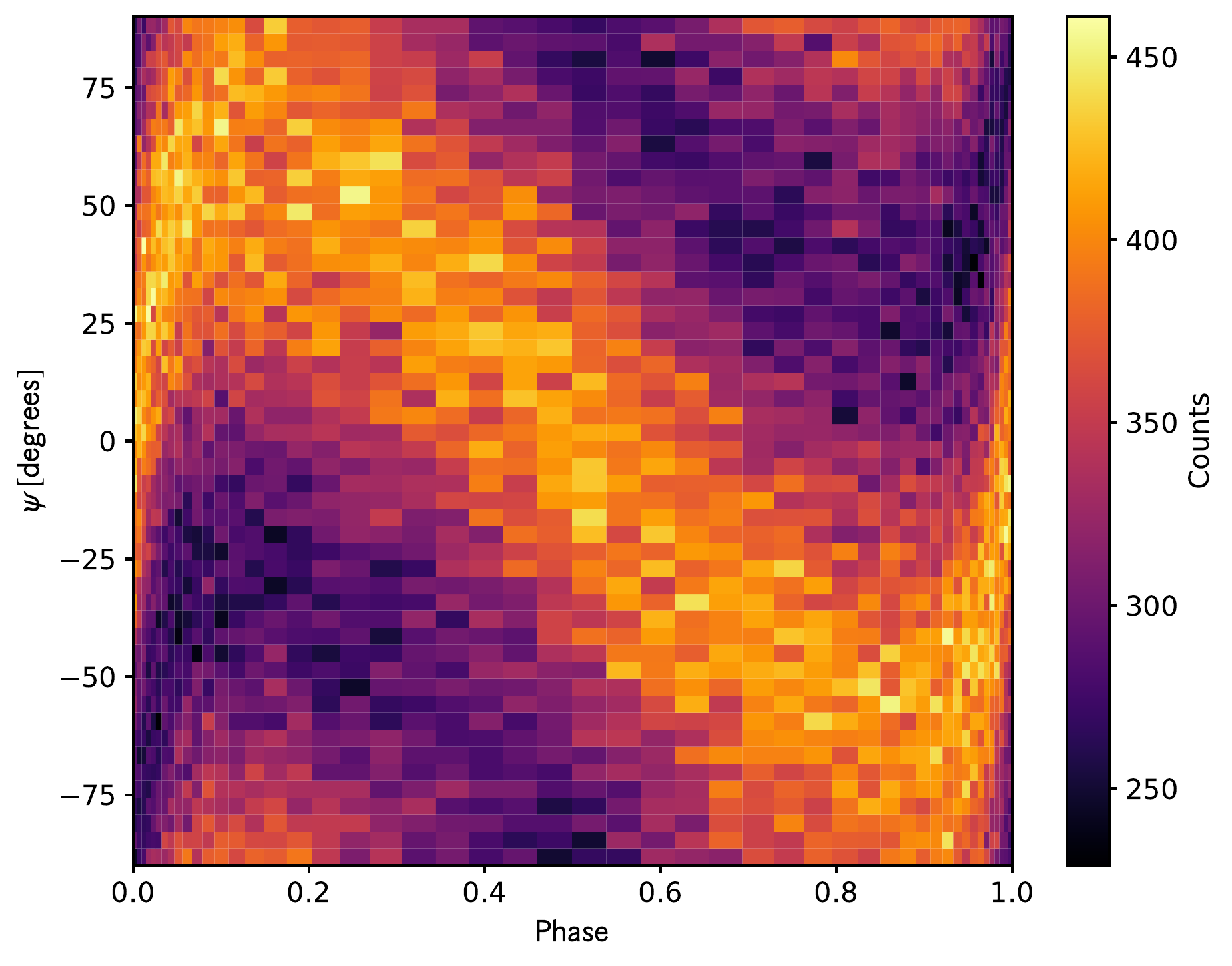}
\caption{\textcolor{black}{2-dimensional, equi-populated histogram   for  the event angle $\psi$  versus the rotational phase of the pulsar ($50\times50$ mesh equi-populated in phase and equi-spaced in angle). The depicted event angles  show collectively, along the main phase-dependent count peak,  a sinusoidal shape as in  Figure\,\ref{fig:geomfit}. However, unlike  the Stokes parameters distribution for ${\bf S_\gamma}$ (see Figure\,\ref{fig:hist2dQU}), which are preferred for performing the unbinned analysis,  the  event angle distribution shows additional secondary count peaks at the top-right and bottom-left corners of the histogram. Events are selected in the $2-8$\,keV range.}}
\label{fig:hist2dang}
\end{figure}

\begin{figure*}
    \centering
    {\Large Binned Analysis}\\
    \includegraphics[width=0.77\textwidth,trim=0 0 2in 0]{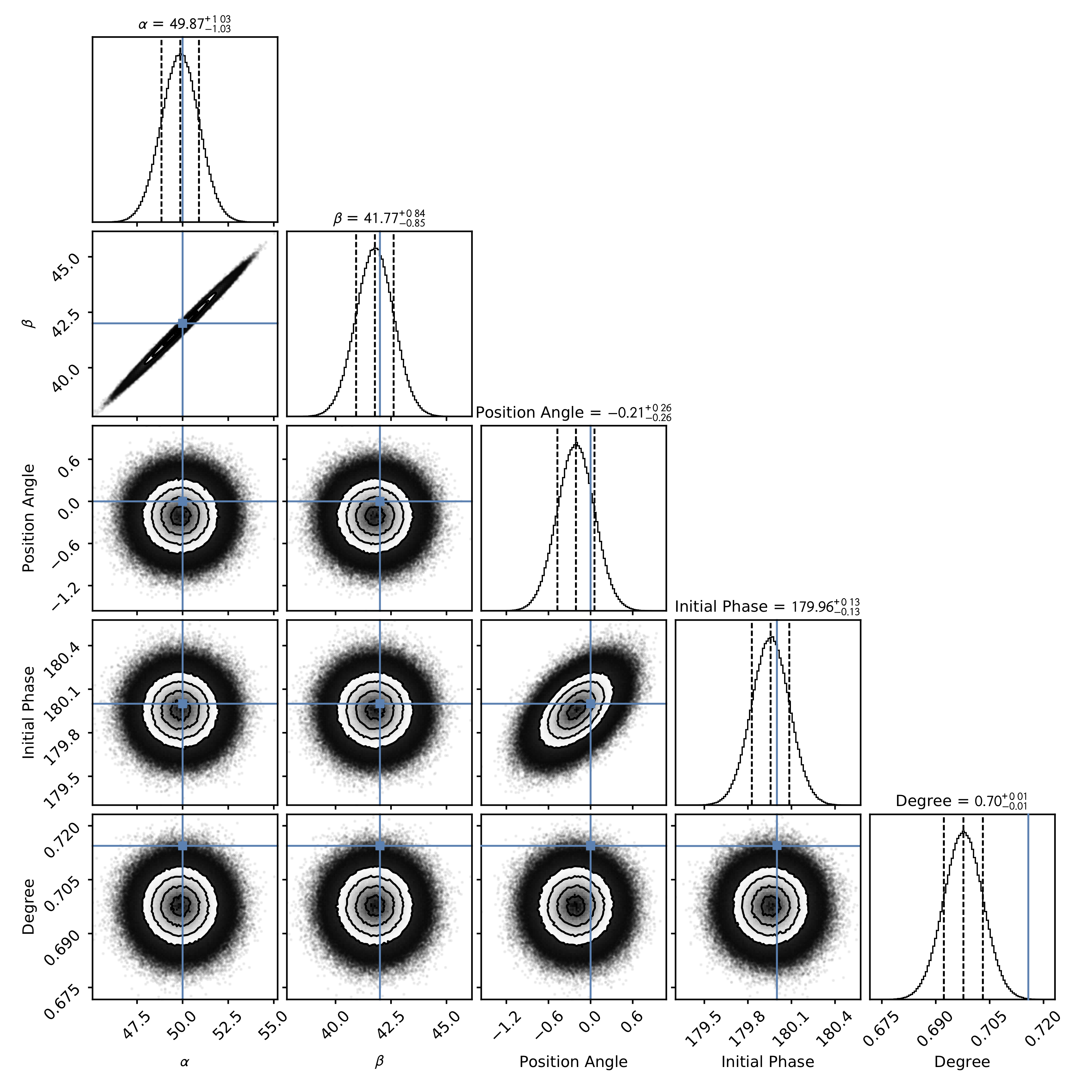}
    \includegraphics[width=0.21\textwidth,trim=7in -3.3in 0 0]{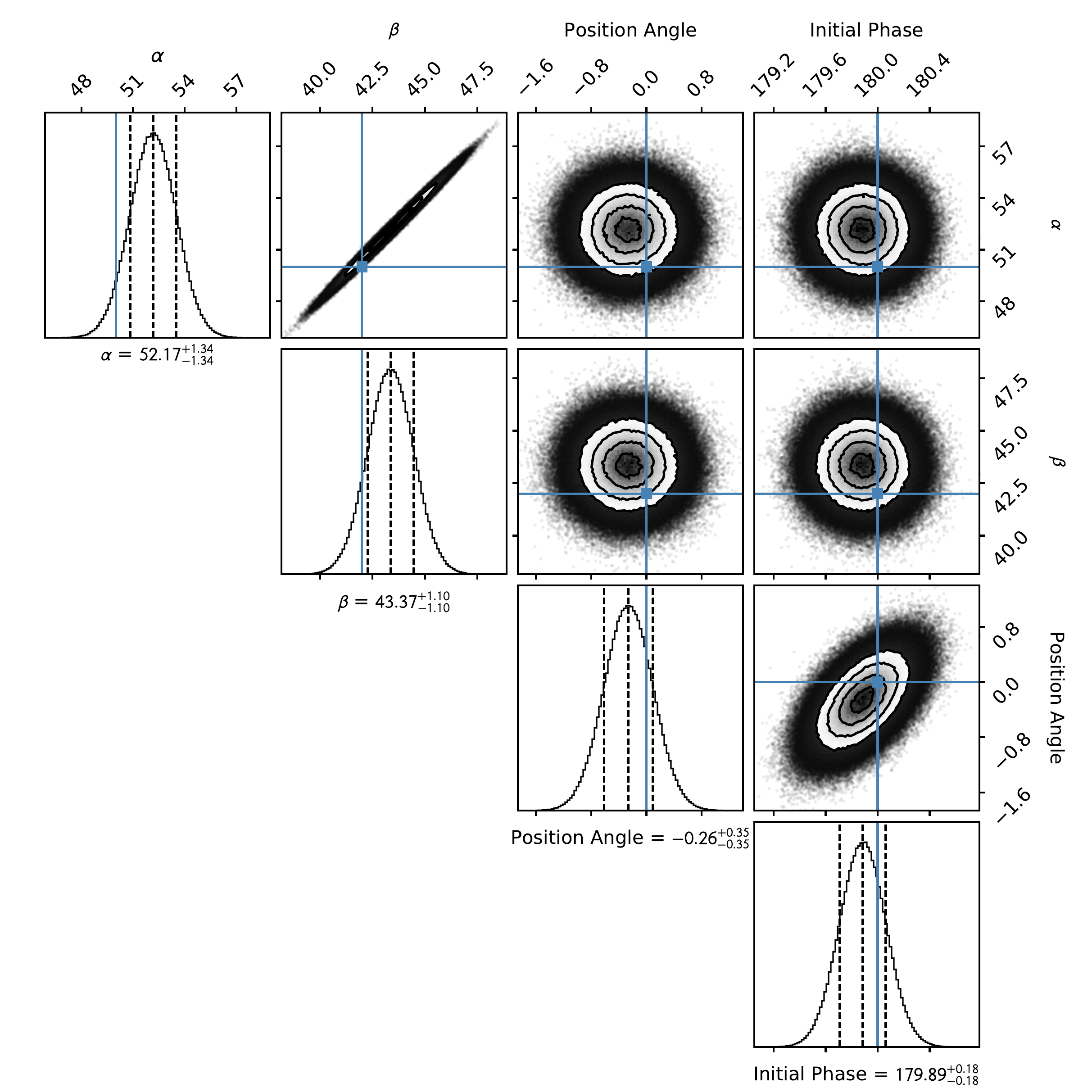}\\
    {\Large Unbinned Analysis}
    \caption{Results of a fit to a single realization of the Her X-1 Observation using the RVM model. The contours depict the uncertainty distribution for a single fit. The blue lines show the true values of the model.}
    \label{fig:geomerr}
\end{figure*}

We compare the RVM to simulated IXPE data of Her X-1 using both binned and unbinned analyses (in the $2-4$ keV range). We first test the methods with a single \texttt{ixpeobssim} realization. We fit the RVM to the polarization angle  considering four free parameters:  $\alpha$ and $\beta$ angles, a rotation in the position angle in the sky and a deviation from the initial phase. Additionally, for the unbinned analysis, we allow the likelihood function to measure the mean polarization degree of the source, leaving $p_0$ as a free parameter that can be adjusted to the data (while fitting the model to the Stokes parameter $Q_\gamma$ and $U_\gamma$ of each event).  Figure\,\ref{fig:geomfit} shows the best fitted RVM curves and Figure \ref{fig:geomerr} shows the associated error parameters obtained with the binned and unbinned analysis \textcolor{black}{(for visualization purpose, Figure\,\ref{fig:hist2dang} also shows a 2-dimensional, equipopulated histogram for the angle $\psi$  associated to each event)}. We found that using  the RVM alone is possible to reconstruct the geometric angles of the pulsar. As shown in Figure\,\ref{fig:geomerr},  the $\alpha$ and $\beta$ solutions obtained with the unbinned technique are well centered around the actual angles of the pulsar. The binned analysis return $\alpha$ and $\beta$ solutions that are slightly biased, a bit more than $1\sigma$ with respect to the actual angles of the input model. We also found that the error parameters obtained with the unbinned technique are smaller than the binned technique. For the single \texttt{ixpeobssim} realization, the errors from the binned technique are approximately  $(\sigma_{\mathrm{bin}}-\sigma_{\mathrm{ubin}})/\sigma_{\mathrm{ubin}} \sim 30\%$ larger than those obtained with the  unbinned  technique.

\begin{figure*}
    \centering
        {\Large Binned Analysis}\\
    \includegraphics[width=0.775\textwidth,trim=0 0 2in 0]{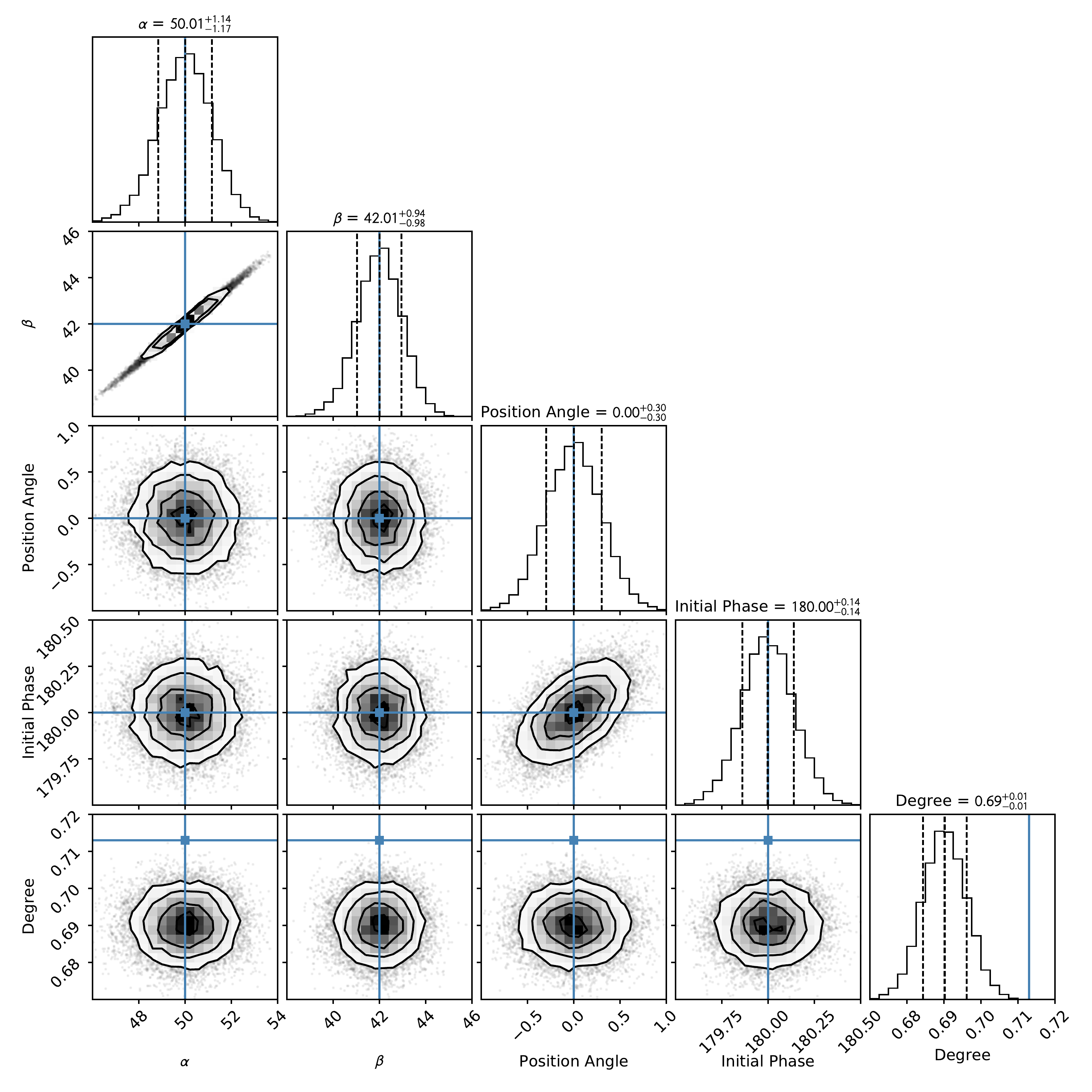}
    \includegraphics[width=0.215\textwidth,trim=7in -3.3in 0 0]{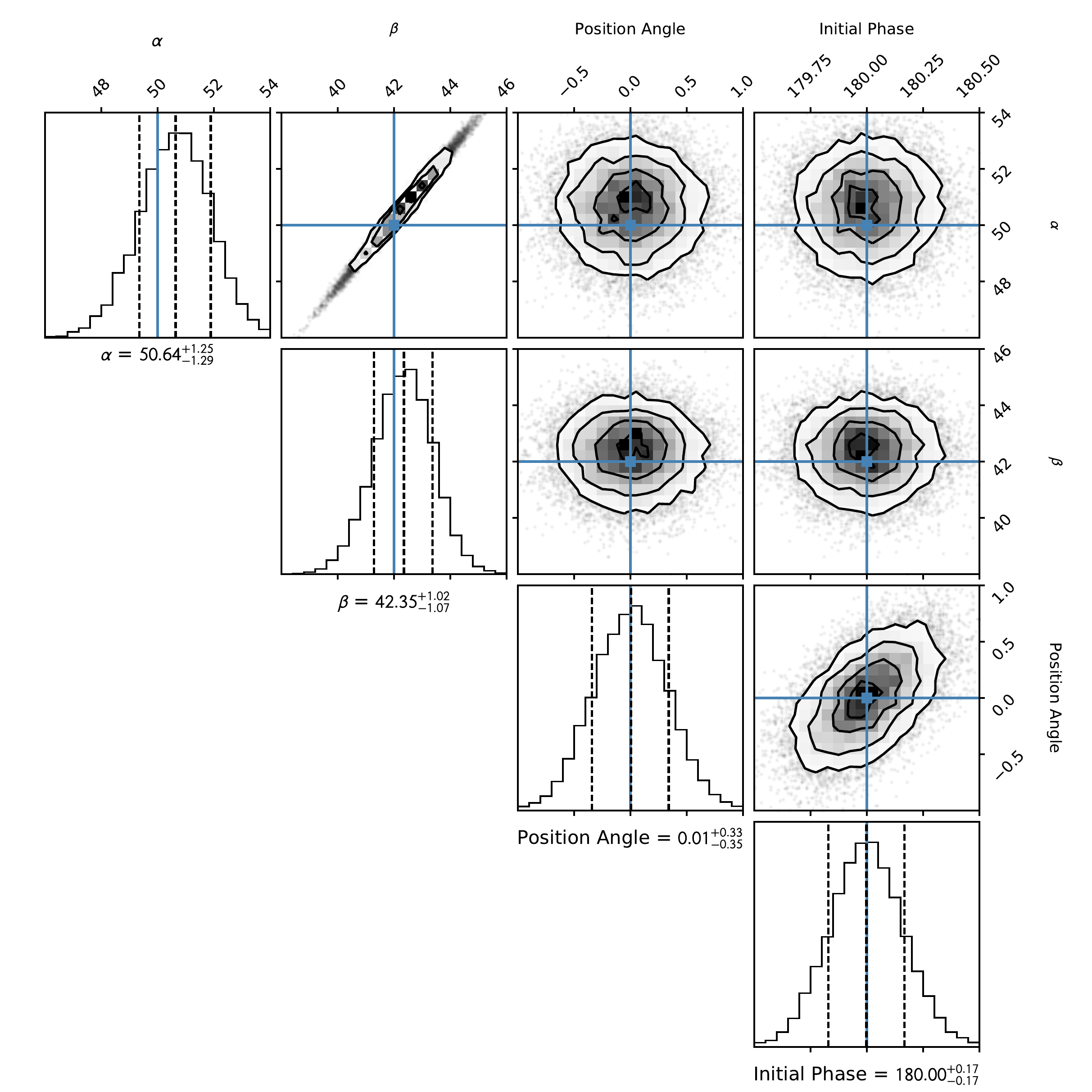} \\
        {\Large Unbinned Analysis}
    \caption{Results of the fit of $12,700$ realizations of the Her X-1 observation using the RVM model. Notice that the unbinned analysis yields an unbiased estimator for the angles $\alpha$ and $\beta$ while the binned approach yields a systematic bias at the one-half-sigma level. The unbinned likelihood function (without energy dispersion) underestimate the mean polarization degree.  }
    \label{fig:geomerr_realizations}
\end{figure*}

Due to the stochastic component of the \texttt{ixpeobssim} simulations, different realizations produce different parameter estimations. In order to further investigate the error estimations and potential bias discussed above, we run multiple \texttt{ixpeobssim} realizations. Figure\,\ref{fig:geomerr_realizations} shows the posterior distribution for the best fitted parameters obtained with $\sim 12,000$ realizations, applying simultaneously the binned and unbinned technique. We found that the binned technique produces error estimations $(\sigma_{\mathrm{bin}}-\sigma_{\mathrm{ubin}})/\sigma_{\mathrm{ubin}} \sim 10\%$ larger than the unbinned technique (which is smaller than the $30\%$ difference found with the single realization). More remarkably, the unbinned technique produces unbiased parameter estimations (except for the mean polarization degree, which is discussed in Sec. \ref{sec:underlying_model} and \ref{sec:energy_dispersion}), while the binned technique produces slightly biased estimations, confirming what we observed with the single realization. In particular, for the $\alpha$ and $\beta$ angles, we found that the bias present in the binned technique is not dramatic and still within the $1\sigma$ error estimation.  

\subsubsection{Underlying model}
\label{sec:underlying_model}
As in previous section, we repeat the polarimetry  analysis, but now comparing the simulated data against the underlying model \citep{2021MNRAS.501..129C}. We proceed with the analysis of multiple \texttt{ixpeobssim} realization as follows:

\begin{itemize}
    \item We consider five free parameters for fitting the model to the data: the $\alpha$ and $\beta$ angles, a rotation of the position angle on the sky, a fractional deviation for the polarization degree with respect to the model and the initial phase when the magnetic axis crosses the meridian of the line of sight.   
    
    \item   In the unbinned technique,  we account only for the detector modulation factor in the likelihood function. 
    
    \item  For the binned technique, the model is binned in phase and energy in the Stokes parameter space, without extra considerations. The best parameter estimation is obtained by minimizing the chi-square  between data and model, accounting for the covariance between $U/I$ and $Q/I$ as explained in \citet{Kislat2015}.
    \item The simulated data for the Stokes $Q$ and $U$ are binned in phase and energy as discussed earlier.
\end{itemize}

Figure \ref{fig:underlying_model_realizations} shows the posterior distribution for the best fitted parameters obtained simultaneously with the  binned and unbinned technique. We confirm the $\sim 10\%$ larger error estimation of the binned technique with respect to those obtained with the unbinned analysis (consistent with the analysis in Sec. \ref{sec:rvm}).
More remarkably, we found a significant bias in the measured polarization degree, more than $1\sigma$ away from the central value of the underlying model (also consistent with the analysis in Sec. \ref{sec:rvm}). That effect is present in both the binned and unbinned analysis, but it is stronger in the latter. \textcolor{black}{The origin of this bias is related to the fact that the polarization degree in the underlying model of \citet{2021MNRAS.501..109C} is energy dependent, decreasing from $\sim75\%$ to $\sim45\%$ in the IXPE band (with an additional dependency on the line-of-sight relative to the magnetic axis). Therefore,  relatively high energies photons, from bands where the emission has low polarization, can contaminate the $2-4$ keV band (and viceversa) due to the energy dispersion of  IXPE detectors, leading to the observed bias.} These results show that, in order to obtain a  precise measurement of the polarization degree of a source, the analysis needs to account for additional instrumental response functions beside the modulation factor, as we discuss in the next section.

\subsection{Unbinned analysis including energy dispersion}
\label{sec:energy_dispersion}
The \texttt{ixpeobssim} package provides additional instrument response functions besides the modulation factor, such as the energy dispersion, effective area,  point spread function, and vignetting of each detector unit. We further refine the unbinned analysis discussed in the previous section by convolving the underlying spectro-polarimetry model with the energy dispersion and effective area, in addition to the modulation factor. The vignetting is not considered in the analysis as it is relevant only for extended sources.

\textcolor{black}{ Following the notation outlined in Sec. \ref{sec:method}, we can describe an event detected by IXPE as ${\bf \hat{S}}_\gamma(E^\prime)$, where $E^\prime$ specifies now an instrumental channel, while the remaining quantities  $M$ and ${\bf S}_m$ depends on the actual energy $E$. We can therefore build an estimator $\propto {\bf \hat{S}}_\gamma(E^\prime) M(E) {\bf S}_m(E) $ that needs to be fully converted to the channel space $E^\prime$.
This can be achieved through a convolution with the instrumental response functions:
$$
\hat{F}_i(E^\prime) \equiv  \frac{1}{2\pi} 
\int {\bf \hat{S}}_\gamma(E^\prime)\,M(E)\, 
{\bf S}_m(E)\, E^{-1}\, A(E)\, R(E,E^\prime)\,  \mathrm{d}E,
$$
where $A$ is the effective area, $R$ is the energy dispersion, and where  $E^{-1}$ is included to convert the intensity-spectrum Stokes model ${\bf S}_m$ to the count-spectrum Stokes model. The latter  is required to properly compute the energy dispersion and count-rate across the instrumental channels, which depends on the count-spectral shape of the source model $C_m = I_m/E$. Next, in order to maximize the impact of the polarization information and minimize the impact of the spectral shape information in the unbinned analysis,  we  build an estimator:
$$\hat{f_i} \equiv \frac{\hat{F}_i}{\hat{C}_m},$$ where 
$$\hat{C}_m(E^\prime) = \int  C_m(E)\, A(E)\, R(E,E^\prime)\, \mathrm{d}E
$$
corresponds to the model counts (per unit time) detected by IXPE.
}

\textcolor{black}{Finally, by expanding  $\hat{f_i}$ and defining more general convolved Stokes models for IXPE detector units $u=1,2,3$: 
\begin{eqnarray*}
\langle \mu_u Q_{m}\rangle \equiv \frac{\int \mu_u(E) Q_m(E,\phi)  C_m(E,\phi) A_u(E) R_u(E,E^\prime) \mathrm{d}E}
{\int  C_m(E,\phi) A_u(E) R_u(E,E^\prime) \mathrm{d}E},&
\\
\langle  \mu_u U_{m}\rangle \equiv \frac{\int  \mu_u(E) U_m(E,\phi) C_m(E,\phi) A_u(E) R_u(E,E^\prime)  \mathrm{d}E}
{\int C_m(E,\phi) A_u(E) R_u(E,E^\prime)   \mathrm{d}E},&
\end{eqnarray*}
we can obtain, in analogy to Sec. \ref{sec:method},  an energy-dispersed estimator to  calculate the  likelihood function:  
\begin{equation*}
\hat{f_i}  =  \frac{1}{2\pi} \left [ 1 + \langle \mu_u Q_{m}\rangle Q_{\gamma} + \langle \mu_u U_{m}\rangle U_{\gamma} \right ].
\end{equation*}
Here, the convolved Stokes models 
$\langle  \mu_u  Q_{m}\rangle$ and $\langle  \mu_u U_{m}\rangle$ 
are phase-dependent,  channel-dependent, and detector-unit-dependent.
However,  with this new likelihood estimator we are not longer able to analyze the polarization information completely alone (as in  in Sec.\,\ref{sec:method}) as it gets mixed with the source spectrum in the convolved models.
}

\begin{figure*}
    \centering
            {\Large Binned Analysis}\\
    \includegraphics[width=0.6\textwidth,trim=0 0 2in 0]{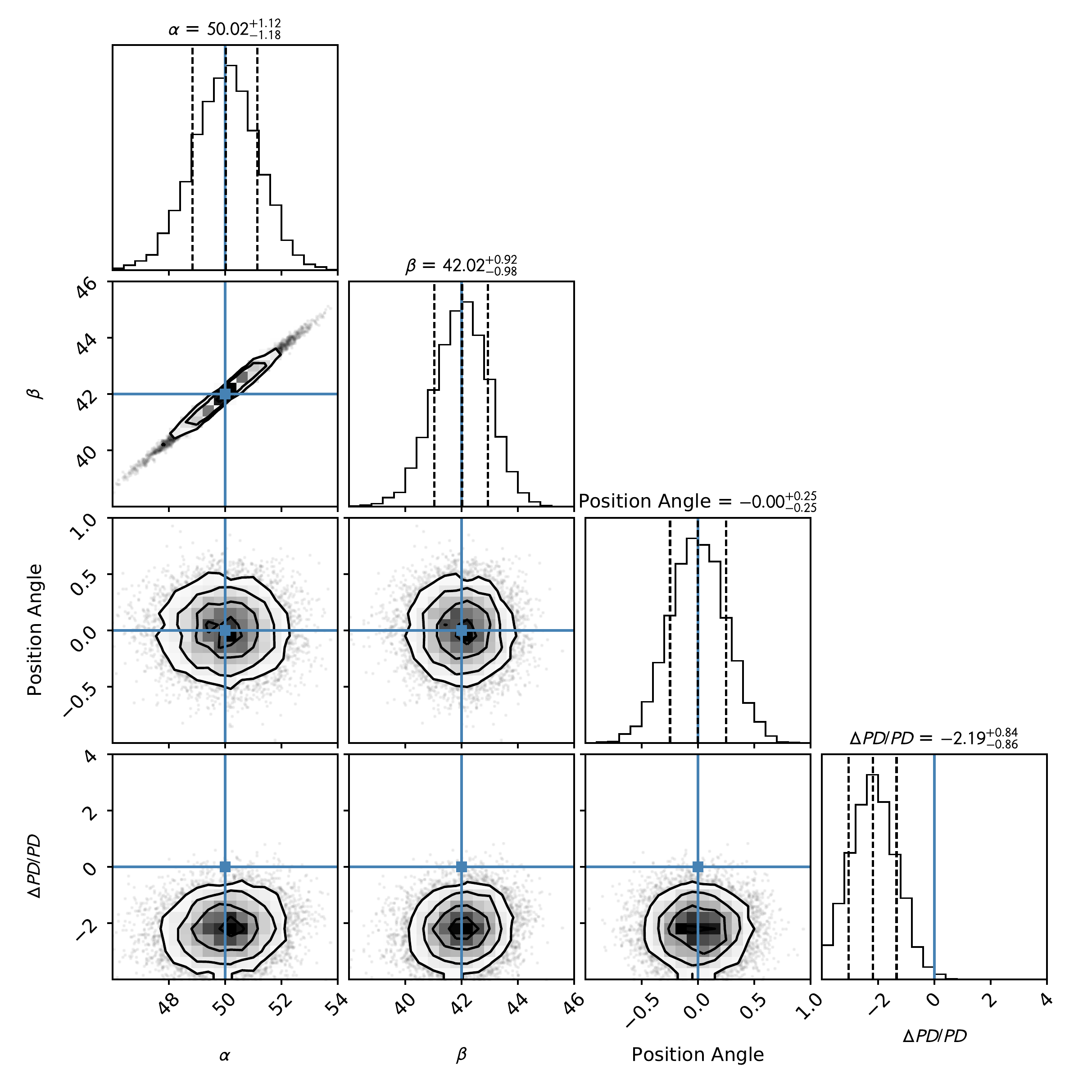}
    \includegraphics[width=0.21\textwidth,trim=7in -3.3in 0 0]{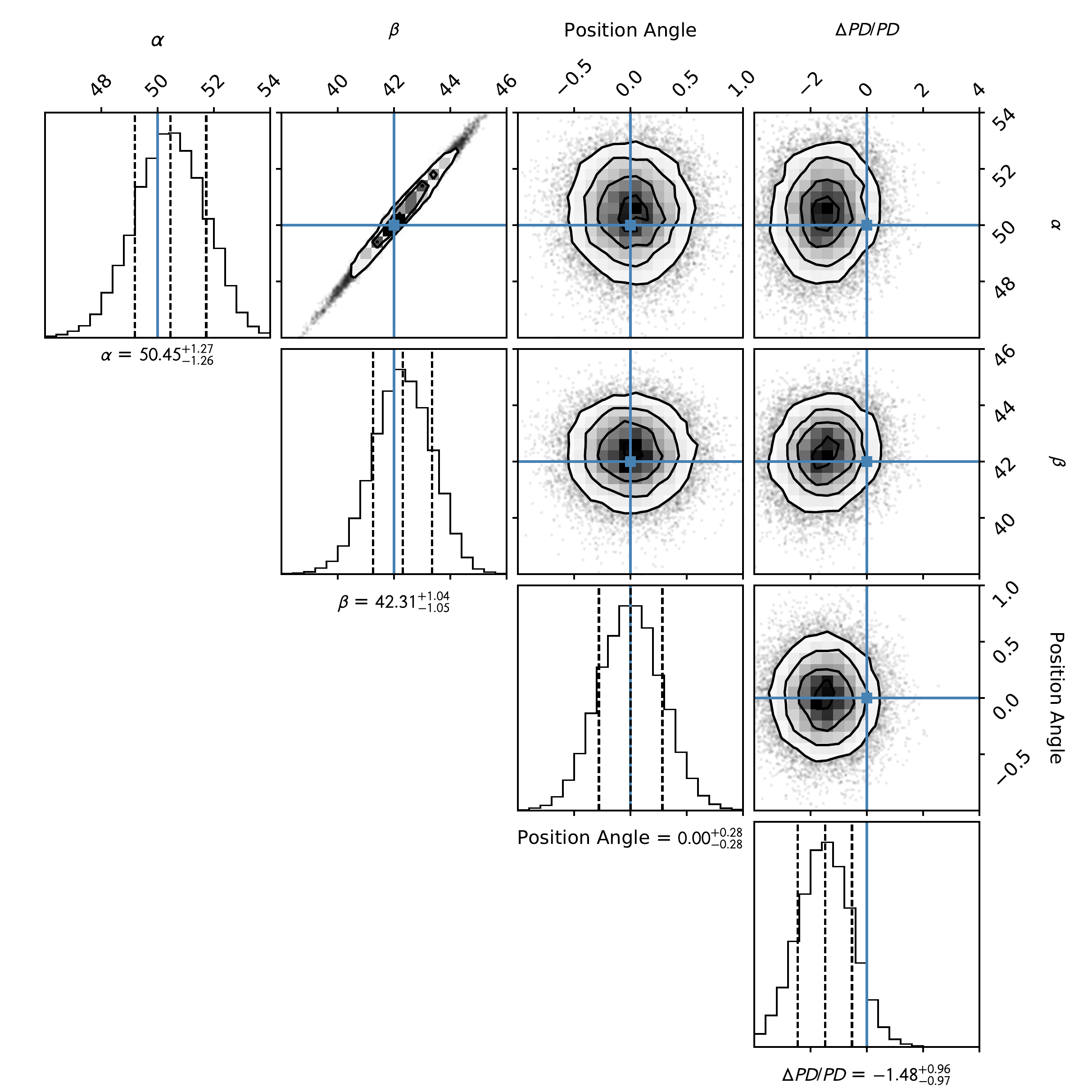}\\
            {\Large Unbinned Analysis}
    \caption{Results of the fit of 7,400 realizations of the Her X-1 Observation using the underlying \citet{2021MNRAS.501..129C} model. Again the unbinned analysis yields an unbiased estimator for the angles $\alpha$ and $\beta$ while the binned approach yields a systematic bias at the one-half-sigma level.  Both techniques underestimate the polarization of the object: they are a biased estimator for the polarization degree, $\Delta PD/PD [\%]$}
    \label{fig:underlying_model_realizations}
\end{figure*}

We repeat the analysis presented in \S~\ref{sec:underlying_model} including now the energy dispersion  (see sample code  in Appendix \ref{sec:appendixA}). 
Figure\,\ref{fig:underlying_model_rmf_realizations} shows the posterior distribution for the best fitted parameters obtained with multiple \texttt{ixpeobssim} realizations. They all remain unchanged for the three parameters $\alpha$, $\beta$, and position angle. However, for the fractional deviation of the polarization degree, the error bars are now centered around zero, removing the bias observed in Figure\,\ref{fig:underlying_model_realizations}. The  polarization degree of the underlying model can be reconstructed with actual precision better than $1\%$. The inclusion of the energy dispersion reduces the bias by a factor of about 24.

\textcolor{black}{By accounting for the instrumental energy dispersion,  we solved the issue of the count contamination of IXPE channels  with photons produced from different energy bands where the emission from the source has different polarization degree. We tested the effect of the energy dispersion for  a source with an intrinsic energy-dependent polarization degree, decreasing from $\sim75\%$ to $\sim 45\%$ in the whole IXPE band (but with energy-independent polarization angle). By ignoring the instrumental energy dispersion, the bias produced in measured parameters might   be even more dramatic if the source also has an intrinsic energy-dependent polarization angle, propagating the bias for example in the measured configuration angles of the source. We leave such study for a future work.}

\section{Conclusions}
\label{sec:conclusions}
We presented a systematic study of the unbinned likelihood technique. We applied it to simulated IXPE observations of Her X-1,  using \texttt{ixpeobssim} package and the  model of \citet{2021MNRAS.501..129C}.
As a first test, we assume minimal knowledge about the physical mechanism responsible for the X-ray polarized emission and analyze the simulated data using the RVM.

We found that the RVM can reconstruct the geometric configuration of Her X-1 using just the phase-dependent polarization angle and the detector modulation factor.
By generating posterior distributions from multiple \texttt{ixpeobssim} realizations, we found the unbinned likelihood technique and the RVM return unbiased configuration angles, while the binned technique returns slightly biased results.

On the other hand, using the underlying model and the modulation factor, we found that both the binned and unbinned likelihood technique return a substantial biased polarization degree relative to the model. Instead, the configuration angles of the system can be reconstructed in a unbiased manner with the unbinned technique, but again they are slightly biased for the binned technique. 

If we account for the energy dispersion of the detector, \textcolor{black}{for which we present an energy-dispersed likelihood estimator}, then the unbinned technique returns an unbiased estimate of the polarization degree. We also show explicitly that the unbinned technique, including the energy dispersion, can be implemented economically, transparently, and in a straightforward manner in \texttt{Python}.  

\section*{Acknowledgements}
We thank the referee for the constructive comments and helpful suggestions. This work was supported by the Natural Sciences and Engineering Research Council of Canada (NSERC), [funding reference \#CITA 490888-16]. IC is a Sherman Fairchild Fellow at Caltech and thanks the Burke Institute at Caltech for supporting her research. J.H. acknowledges support from the Canadian Space Agency through the co-investigator grant program and
from the NSERC through the Discovery Grant Program.
This research was enabled in part by support provided by WestGrid (www.westgrid.ca) and Compute Canada (www.computecanada.ca).
This research makes use of the SciServer science platform (www.sciserver.org) and UBC ARC Sockeye infrastructure.

\begin{figure*}
    \centering
    \includegraphics[width=0.8\textwidth,trim=0 0 0in 0]{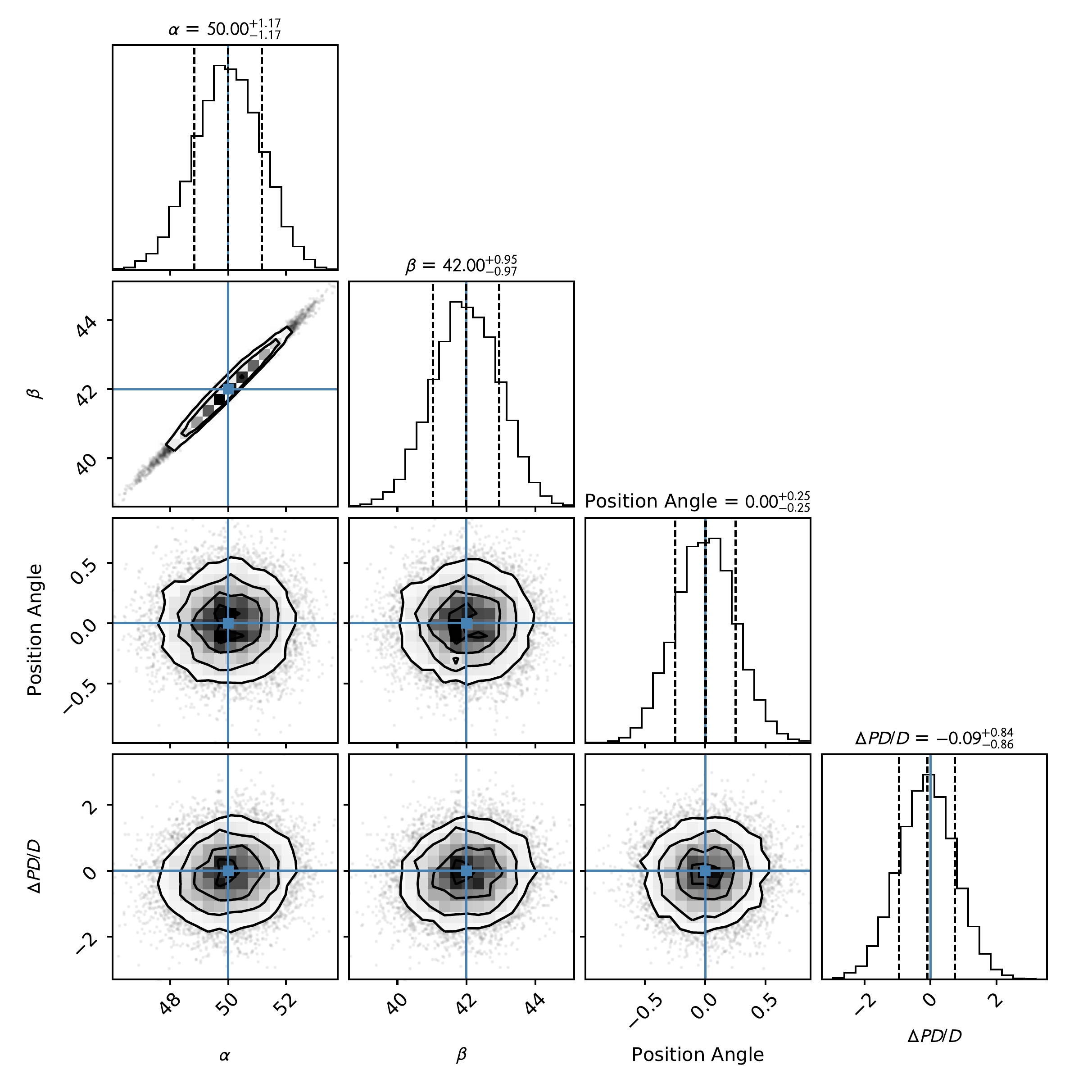}
    \caption{Results of the fit of 9,400 realizations of the Her X-1 Observation using the underlying \citet{2021MNRAS.501..129C} model including energy dispersion.  All of the values are now unbiased with typically uncertainties of one percent including $\Delta PD/PD [\%]$.
  }
    \label{fig:underlying_model_rmf_realizations}
\end{figure*}

\section*{Data Availability}
A Jupyter Notebook with the main calculations is available in the link below:\\ \url{https://github.com/UBC-Astrophysics/Unbinned-Xray-polarimetry}
 



\bibliographystyle{mnras}
\bibliography{main} 

\begin{thebibliography}{}
\makeatletter
\relax
\def\mn@urlcharsother{\let\do\@makeother \do\$\do\&\do\#\do\^\do\_\do\%\do\~}
\def\mn@doi{\begingroup\mn@urlcharsother \@ifnextchar [ {\mn@doi@}
  {\mn@doi@[]}}
\def\mn@doi@[#1]#2{\def\@tempa{#1}\ifx\@tempa\@empty \href
  {http://dx.doi.org/#2} {doi:#2}\else \href {http://dx.doi.org/#2} {#1}\fi
  \endgroup}
\def\mn@eprint#1#2{\mn@eprint@#1:#2::\@nil}
\def\mn@eprint@arXiv#1{\href {http://arxiv.org/abs/#1} {{\tt arXiv:#1}}}
\def\mn@eprint@dblp#1{\href {http://dblp.uni-trier.de/rec/bibtex/#1.xml}
  {dblp:#1}}
\def\mn@eprint@#1:#2:#3:#4\@nil{\def\@tempa {#1}\def\@tempb {#2}\def\@tempc
  {#3}\ifx \@tempc \@empty \let \@tempc \@tempb \let \@tempb \@tempa \fi \ifx
  \@tempb \@empty \def\@tempb {arXiv}\fi \@ifundefined
  {mn@eprint@\@tempb}{\@tempb:\@tempc}{\expandafter \expandafter \csname
  mn@eprint@\@tempb\endcsname \expandafter{\@tempc}}}

\bibitem[\protect\citeauthoryear{{Baldini} et~al.,}{{Baldini}
  et~al.}{2021}]{Baldini2021}
{Baldini} L.,  et~al., 2021, \mn@doi [Astroparticle Physics]
  {10.1016/j.astropartphys.2021.102628}, \href
  {https://ui.adsabs.harvard.edu/abs/2021APh...13302628B} {133, 102628}

\bibitem[\protect\citeauthoryear{{Baldini}, {Bucciantini}, {Di Lalla},
  {Ehlert}, {Manfreda}, {Omodei}, {Pesce-Rollins}  \& {Sgr{\`o}}}{{Baldini}
  et~al.}{2022}]{Baldini2022}
{Baldini} L.,  {Bucciantini} N.,  {Di Lalla} N.,  {Ehlert} S.~R.,  {Manfreda}
  A.,  {Omodei} N.,  {Pesce-Rollins} M.,   {Sgr{\`o}} C.,  2022, arXiv
  e-prints, \href {https://ui.adsabs.harvard.edu/abs/2022arXiv220306384B} {p.
  arXiv:2203.06384}

\bibitem[\protect\citeauthoryear{{Beilicke} et~al.,}{{Beilicke}
  et~al.}{2014}]{Beilicke2014}
{Beilicke} M.,  et~al., 2014, \mn@doi [Journal of Astronomical Instrumentation]
  {10.1142/S225117171440008X}, \href
  {https://ui.adsabs.harvard.edu/abs/2014JAI.....340008B} {3, 1440008}

\bibitem[\protect\citeauthoryear{{Bellazzini} et~al.,}{{Bellazzini}
  et~al.}{2007}]{Bellazzini2007}
{Bellazzini} R.,  et~al., 2007, \mn@doi [Nuclear Instruments and Methods in
  Physics Research A] {10.1016/j.nima.2007.05.304}, \href
  {https://ui.adsabs.harvard.edu/abs/2007NIMPA.579..853B} {579, 853}

\bibitem[\protect\citeauthoryear{{Caiazzo} \& {Heyl}}{{Caiazzo} \&
  {Heyl}}{2021a}]{2021MNRAS.501..109C}
{Caiazzo} I.,  {Heyl} J.,  2021a, \mn@doi [\mnras] {10.1093/mnras/staa3428},
  \href {https://ui.adsabs.harvard.edu/abs/2021MNRAS.501..109C} {501, 109}

\bibitem[\protect\citeauthoryear{{Caiazzo} \& {Heyl}}{{Caiazzo} \&
  {Heyl}}{2021b}]{2021MNRAS.501..129C}
{Caiazzo} I.,  {Heyl} J.,  2021b, \mn@doi [\mnras] {10.1093/mnras/staa3429},
  \href {https://ui.adsabs.harvard.edu/abs/2021MNRAS.501..129C} {501, 129}

\bibitem[\protect\citeauthoryear{{Chauvin} et~al.,}{{Chauvin}
  et~al.}{2018}]{Chauvin2018}
{Chauvin} M.,  et~al., 2018, \mn@doi [\mnras] {10.1093/mnrasl/sly027}, \href
  {https://ui.adsabs.harvard.edu/abs/2018MNRAS.477L..45C} {477, L45}

\bibitem[\protect\citeauthoryear{{Costa}, {Soffitta}, {Bellazzini}, {Brez},
  {Lumb}  \& {Spandre}}{{Costa} et~al.}{2001}]{Costa2001}
{Costa} E.,  {Soffitta} P.,  {Bellazzini} R.,  {Brez} A.,  {Lumb} N.,
  {Spandre} G.,  2001, \nat, \href
  {https://ui.adsabs.harvard.edu/abs/2001Natur.411..662C} {411, 662}

\bibitem[\protect\citeauthoryear{{Feng} et~al.,}{{Feng} et~al.}{2019}]{Feng19}
{Feng} H.,  et~al., 2019, \mn@doi [Experimental Astronomy]
  {10.1007/s10686-019-09625-z}, \href
  {https://ui.adsabs.harvard.edu/abs/2019ExA....47..225F} {47, 225}

\bibitem[\protect\citeauthoryear{{F{\"u}rst} et~al.,}{{F{\"u}rst}
  et~al.}{2013}]{Furst2013}
{F{\"u}rst} F.,  et~al., 2013, \mn@doi [\apj] {10.1088/0004-637X/779/1/69},
  \href {https://ui.adsabs.harvard.edu/abs/2013ApJ...779...69F} {779, 69}

\bibitem[\protect\citeauthoryear{{Heisenberg} \& {Euler}}{{Heisenberg} \&
  {Euler}}{1936}]{Heisenberg1936}
{Heisenberg} W.,  {Euler} H.,  1936, \mn@doi [Zeitschrift fur Physik]
  {10.1007/BF01343663}, \href
  {https://ui.adsabs.harvard.edu/abs/1936ZPhy...98..714H} {98, 714}

\bibitem[\protect\citeauthoryear{{Heyl} \& {Shaviv}}{{Heyl} \&
  {Shaviv}}{2000}]{Heyl2000}
{Heyl} J.~S.,  {Shaviv} N.~J.,  2000, \mn@doi [\mnras]
  {10.1046/j.1365-8711.2000.03076.x}, \href
  {https://ui.adsabs.harvard.edu/abs/2000MNRAS.311..555H} {311, 555}

\bibitem[\protect\citeauthoryear{{Heyl} \& {Shaviv}}{{Heyl} \&
  {Shaviv}}{2002}]{Heyl2002}
{Heyl} J.~S.,  {Shaviv} N.~J.,  2002, \mn@doi [\prd]
  {10.1103/PhysRevD.66.023002}, \href
  {https://ui.adsabs.harvard.edu/abs/2002PhRvD..66b3002H} {66, 023002}

\bibitem[\protect\citeauthoryear{{Jahoda} et~al.,}{{Jahoda}
  et~al.}{2019}]{Jahoda19}
{Jahoda} K.,  et~al., 2019, arXiv e-prints, \href
  {https://ui.adsabs.harvard.edu/abs/2019arXiv190710190J} {p. arXiv:1907.10190}

\bibitem[\protect\citeauthoryear{{Kaspi} \& {Beloborodov}}{{Kaspi} \&
  {Beloborodov}}{2017}]{Kaspi17}
{Kaspi} V.~M.,  {Beloborodov} A.~M.,  2017, \mn@doi [\araa]
  {10.1146/annurev-astro-081915-023329}, \href
  {https://ui.adsabs.harvard.edu/abs/2017ARA&A..55..261K} {55, 261}

\bibitem[\protect\citeauthoryear{{Kislat}, {Clark}, {Beilicke}  \&
  {Krawczynski}}{{Kislat} et~al.}{2015}]{Kislat2015}
{Kislat} F.,  {Clark} B.,  {Beilicke} M.,   {Krawczynski} H.,  2015, \mn@doi
  [Astroparticle Physics] {10.1016/j.astropartphys.2015.02.007}, \href
  {https://ui.adsabs.harvard.edu/abs/2015APh....68...45K} {68, 45}

\bibitem[\protect\citeauthoryear{{Marshall}}{{Marshall}}{2021}]{2021AJ....162..134M}
{Marshall} H.~L.,  2021, \mn@doi [\aj] {10.3847/1538-3881/ac173d}, \href
  {https://ui.adsabs.harvard.edu/abs/2021AJ....162..134M} {162, 134}

\bibitem[\protect\citeauthoryear{{Marshall} et~al.,}{{Marshall}
  et~al.}{2021}]{Marshall2021}
{Marshall} H.~L.,  et~al., 2021, in Society of Photo-Optical Instrumentation
  Engineers (SPIE) Conference Series. p. 118220O, \mn@doi{10.1117/12.2596186}

\bibitem[\protect\citeauthoryear{{Meszaros}}{{Meszaros}}{1992}]{Meszaros92}
{Meszaros} P.,  1992, {High-energy radiation from magnetized neutron stars}

\bibitem[\protect\citeauthoryear{{Peirson} \& {Romani}}{{Peirson} \&
  {Romani}}{2021}]{Peirson2021b}
{Peirson} A.~L.,  {Romani} R.~W.,  2021, \mn@doi [\apj]
  {10.3847/1538-4357/ac157d}, \href
  {https://ui.adsabs.harvard.edu/abs/2021ApJ...920...40P} {920, 40}

\bibitem[\protect\citeauthoryear{{Peirson}, {Romani}, {Marshall}, {Steiner}  \&
  {Baldini}}{{Peirson} et~al.}{2021}]{Peirson2021a}
{Peirson} A.~L.,  {Romani} R.~W.,  {Marshall} H.~L.,  {Steiner} J.~F.,
  {Baldini} L.,  2021, \mn@doi [Nuclear Instruments and Methods in Physics
  Research A] {10.1016/j.nima.2020.164740}, \href
  {https://ui.adsabs.harvard.edu/abs/2021NIMPA.98664740P} {986, 164740}

\bibitem[\protect\citeauthoryear{{Pesce-Rollins}, {Lalla}, {Omodei}  \&
  {Baldini}}{{Pesce-Rollins} et~al.}{2019}]{Pesce2019}
{Pesce-Rollins} M.,  {Lalla} N.~D.,  {Omodei} N.,   {Baldini} L.,  2019,
  \mn@doi [Nuclear Instruments and Methods in Physics Research A]
  {10.1016/j.nima.2018.10.041}, \href
  {https://ui.adsabs.harvard.edu/abs/2019NIMPA.936..224P} {936, 224}

\bibitem[\protect\citeauthoryear{{Radhakrishnan} \& {Cooke}}{{Radhakrishnan} \&
  {Cooke}}{1969}]{Radhakrishnan1969}
{Radhakrishnan} V.,  {Cooke} D.~J.,  1969, \aplett, \href
  {https://ui.adsabs.harvard.edu/abs/1969ApL.....3..225R} {3, 225}

\bibitem[\protect\citeauthoryear{{Rankin} et~al.,}{{Rankin}
  et~al.}{2022}]{Rankin22}
{Rankin} J.,  et~al., 2022, \mn@doi [\aj] {10.3847/1538-3881/ac397f}, \href
  {https://ui.adsabs.harvard.edu/abs/2022AJ....163...39R} {163, 39}

\bibitem[\protect\citeauthoryear{{Schwinger}}{{Schwinger}}{1951}]{Schwinger1951}
{Schwinger} J.,  1951, \mn@doi [Physical Review] {10.1103/PhysRev.82.664},
  \href {https://ui.adsabs.harvard.edu/abs/1951PhRv...82..664S} {82, 664}

\bibitem[\protect\citeauthoryear{{Sgr{\`o}}}{{Sgr{\`o}}}{2017}]{2017SPIE10397E..0FS}
{Sgr{\`o}} C.,  2017, in Society of Photo-Optical Instrumentation Engineers
  (SPIE) Conference Series. p. 103970F, \mn@doi{10.1117/12.2273922}

\bibitem[\protect\citeauthoryear{{She} et~al.,}{{She} et~al.}{2015}]{She2015}
{She} R.,  et~al., 2015, in {Siegmund} O.~H.,  ed.,  Society of Photo-Optical
  Instrumentation Engineers (SPIE) Conference Series Vol. 9601, UV, X-Ray, and
  Gamma-Ray Space Instrumentation for Astronomy XIX. p. 96010I (\mn@eprint
  {arXiv} {1509.04392}), \mn@doi{10.1117/12.2186133}

\bibitem[\protect\citeauthoryear{{Soffitta} et~al.,}{{Soffitta}
  et~al.}{2021}]{Soffitta2021}
{Soffitta} P.,  et~al., 2021, \mn@doi [\aj] {10.3847/1538-3881/ac19b0}, \href
  {https://ui.adsabs.harvard.edu/abs/2021AJ....162..208S} {162, 208}

\bibitem[\protect\citeauthoryear{{Thompson}, {Lyutikov}  \&
  {Kulkarni}}{{Thompson} et~al.}{2002}]{Thompson02}
{Thompson} C.,  {Lyutikov} M.,   {Kulkarni} S.~R.,  2002, \mn@doi [\apj]
  {10.1086/340586}, \href
  {https://ui.adsabs.harvard.edu/abs/2002ApJ...574..332T} {574, 332}

\bibitem[\protect\citeauthoryear{{Turolla}, {Zane}  \& {Watts}}{{Turolla}
  et~al.}{2015}]{Turolla15}
{Turolla} R.,  {Zane} S.,   {Watts} A.~L.,  2015, \mn@doi [Reports on Progress
  in Physics] {10.1088/0034-4885/78/11/116901}, \href
  {https://ui.adsabs.harvard.edu/abs/2015RPPh...78k6901T} {78, 116901}

\bibitem[\protect\citeauthoryear{{Vadawale}, {Paul}, {Pendharkar}  \&
  {Naik}}{{Vadawale} et~al.}{2010}]{Vadawale2010}
{Vadawale} S.~V.,  {Paul} B.,  {Pendharkar} J.,   {Naik} S.,  2010, \mn@doi
  [Nuclear Instruments and Methods in Physics Research A]
  {10.1016/j.nima.2010.02.116}, \href
  {https://ui.adsabs.harvard.edu/abs/2010NIMPA.618..182V} {618, 182}

\bibitem[\protect\citeauthoryear{{Weisskopf}}{{Weisskopf}}{1936}]{Weisskopf1936}
{Weisskopf} V.,  1936, {Kongelige Danske Videnskabernes Selskab,
  Matematisk-fysiske Meddelelser}, 14, 714

\bibitem[\protect\citeauthoryear{{Weisskopf} et~al.,}{{Weisskopf}
  et~al.}{2016}]{Weisskopf16}
{Weisskopf} M.~C.,  et~al., 2016, in {den Herder} J.-W.~A.,  {Takahashi} T.,
  {Bautz} M.,  eds,  Society of Photo-Optical Instrumentation Engineers (SPIE)
  Conference Series Vol. 9905, Space Telescopes and Instrumentation 2016:
  Ultraviolet to Gamma Ray. p. 990517, \mn@doi{10.1117/12.2235240}

\bibitem[\protect\citeauthoryear{{Weisskopf} et~al.}{{Weisskopf}
  et~al.}{2021}]{2021arXiv211201269W}
{Weisskopf} M.~C.,  et~al., 2021, arXiv e-prints, \href
  {https://ui.adsabs.harvard.edu/abs/2021arXiv211201269W} {p. arXiv:2112.01269}

\bibitem[\protect\citeauthoryear{{Wolff} et~al.,}{{Wolff}
  et~al.}{2016}]{Wolff2016}
{Wolff} M.~T.,  et~al., 2016, \mn@doi [\apj] {10.3847/0004-637X/831/2/194},
  \href {https://ui.adsabs.harvard.edu/abs/2016ApJ...831..194W} {831, 194}

\bibitem[\protect\citeauthoryear{{Zhang} et~al.,}{{Zhang}
  et~al.}{2016}]{Zhang16}
{Zhang} S.~N.,  et~al., 2016, in {den Herder} J.-W.~A.,  {Takahashi} T.,
  {Bautz} M.,  eds,  Society of Photo-Optical Instrumentation Engineers (SPIE)
  Conference Series Vol. 9905, Space Telescopes and Instrumentation 2016:
  Ultraviolet to Gamma Ray. p. 99051Q (\mn@eprint {arXiv} {1607.08823}),
  \mn@doi{10.1117/12.2232034}

\makeatother
\end{thebibliography}



\clearpage
\appendix
\onecolumn
\section{Sample Code for Likelihood Calculations}
\label{sec:appendixA}
\begin{figure}
    \centering
    \includegraphics[width=\columnwidth]{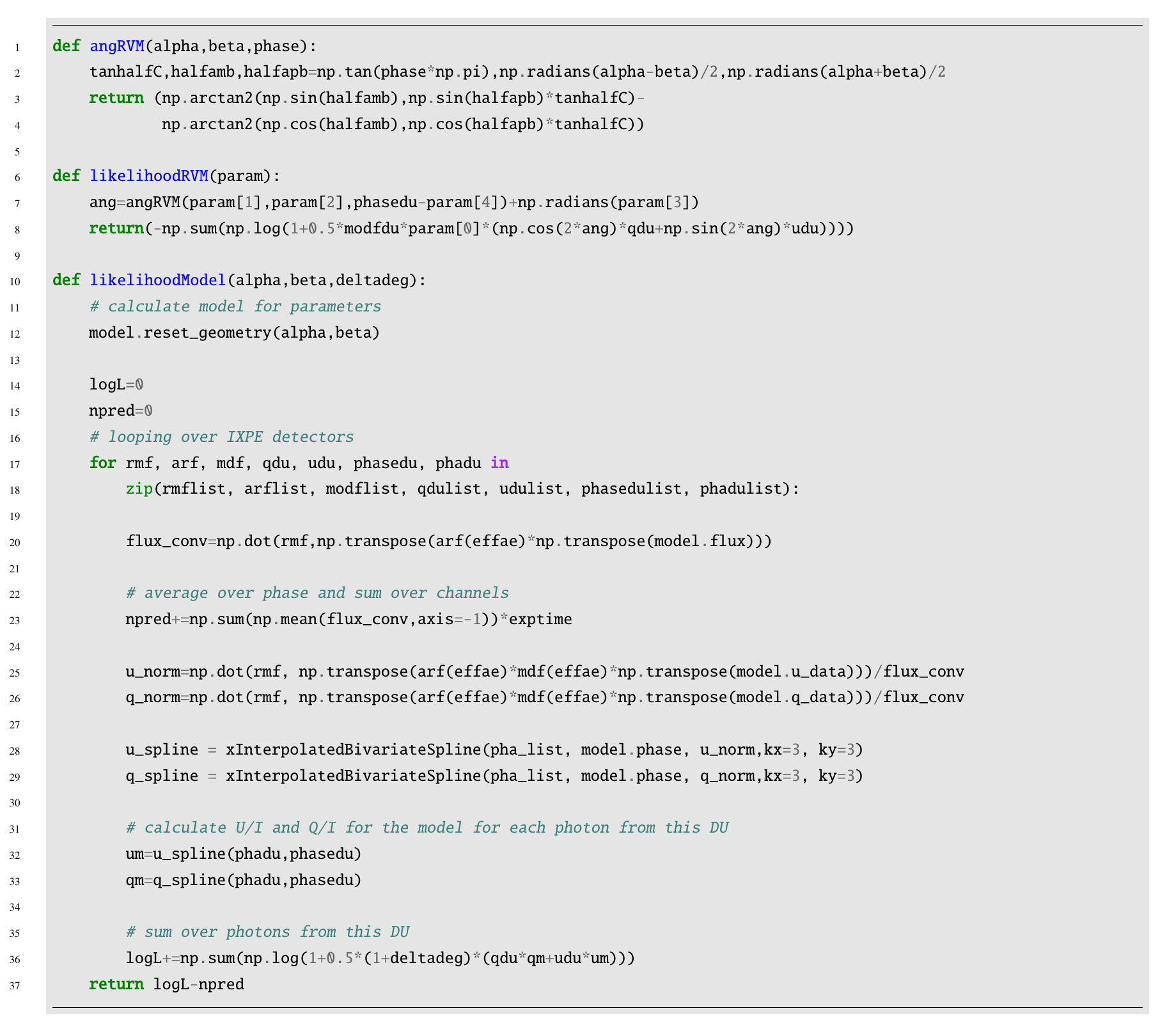}
    \caption{Python Code for RVM Model and General Model including Energy Dispersion.}
    \label{fig:code}
\end{figure}

A code snippet of the unbinned technique including  the energy dispersion is shown in Figure \ref{fig:code}. 
 The likelihood function for the   underlying model is defined in the function \texttt{likelihoodModel}. The main \texttt{Python `for'} loop iterates over the 3 detector units of IXPE. The instruments response functions \texttt{rmf}, \texttt{arf}, and \texttt{mdf} correspond to the energy dispersion (loaded as a matrix), effective area (loaded as a function), and the modulation factor (loaded as a function), respectively,  which are obtained from \texttt{ixpeobssim}. The event information \texttt{qdu}, \texttt{udu}, \texttt{phasedu},   \texttt{phadu} correspond to the Stokes  $q$ and $u$, phase, and event pulse height, respectively. The \texttt{`model'} class loads the Her X-1 model for the flux and Stokes $Q$ and $U$, which are energy- and phase-dependent, as well as geometry-dependent on the $\alpha$ and $\beta$ angles. The convolved models $\langle \mu_u Q_{m}\rangle$ and $\langle \mu_u U_{m}\rangle$ are computed in \texttt{qm} and  \texttt{um}.

\bsp	
\label{lastpage}
\end{document}